\documentclass[a4eurescofig1.epspaper,12pt,fleqn]{article}             
\usepackage{epsf}
\usepackage{fancybox}
\usepackage[dvips]{pstcol}
\usepackage{graphicx}
\usepackage{psfrag}
\usepackage{amssymb}
\usepackage{latexsym}
\usepackage{amsmath}
\usepackage{amscd}
\def\BT{\begin{tabular}}
\def\ET{\end{tabular}}
\def\BE{\begin{eqnarray}}
\def\EE{\end{eqnarray}}
\begin{document}

\vspace*{2.0cm}
{\Large \bf Translational and Rotational Diffusion
in Stretched Water}\\

P. A. Netz$^{a}$, F. Starr $^b$,
M. C. Barbosa$^c$ e H.~Eugene Stanley$^d$\\

$^a$Departamento  de Qu\'{\i}mica, ULBRA,
Canoas RS, BRASIL; 

$^b$Center for Theoretical and Computational Materials 
Science and Polymers Division, National Institute
of Standards and Technology, Gaithersburg, 
Maryland 20899, USA;

$^c$Departamento  de F\'{\i}sica, UFRGS,
Porto Alegre RS, BRASIL; 

$^d$Center of Polymer Studies - Boston University
Boston, MA 02215, USA

\vskip 1.0\baselineskip

We perform molecular dynamics simulations using the 
extended simple point charge SPC/E water model
in order to investigate the dynamical behavior of 
supercooled-stretched water. We focus on the 
behavior of the translational diffusion coefficient,
the orientational relaxation time, and the local 
hydrogen bond network. Decreasing density or pressure
along an isothermal path, there is a mininum in the 
diffusion coefficient and a maximum in the orientational
relaxation time, suggesting an initial enhancement
and subsequent breakdown of the tetrahedral structure 
and of the hydrogen bond network as the
density decreases. The analysis of the tetrahedral 
structure of the nearest neighbors help to clarify
the relationship between the local structural changes
and the system dynamics. We also find that the product
of diffusion coefficient and relaxation time is nearly
constant, indicating that the anomalous behavior observed
in the rotational and translational diffusion
arise from the same microscopic mechanism.\\

\section{Introduction}

Water is one of the most intriguing subjects of research from the 
physical, chemical or biological point of view.
Due to the ubiquity and  anomalous thermodynamic  behavior of water,
it has  been studied for centuries,
and even now unexpected phenomena  
continue to be discovered~\cite{deb96}.\\  

Ordinary liquid water is highly locally ordered, with
each molecule having almost exactly four tetrahedrally coordinated
neighbors forming a connected random network of
hydrogen bonds well above the percolation threshold.
Despite possible expectations that 
water might behave like
a stable network, the hydrogen bond network restructures
itself in picosecond time scale. This
indicates that the random tetrahedral
network is not perfect, but  contains some 
sort of structural defects, what it is supported by
experiments.
The local structure is sensitive to variable temperature or 
pressure, which can result in modifications such as 
distortion of the hydrogen bond angles
and distances, as well as increase (or decrease) in the number
of neighbors. These changes in local order have been shown to be
intimately connected with the thermodynamic anomalies~\cite{chaplin}.
This is the case of the existence of a maximum in the density at
4$^o$C, unusually high melting, boiling and critical points,
existence of a minimum in the constant pressure heat 
capacity, among several others.\\

These local network defects are also important for
the mobility of water. In contrast to conventional wisdom,
the increase of local density (and therefore the increase
in pressure) leads to increased mobility in ambient water,
a result of weakening and possible destructions of hydrogen
bonds that favor an open structure, as shown by
experimental observations\cite{jon76,pri87}
and simulations~\cite{sta99,bag97,err01}.  
At sufficiently large pressure, steric constraints become
significant and mobility decreases on increasing
density, as in common liquids.  As a result,
along each isotherm, there is a maximum of the diffusion coefficient
as a function of density.\\

If the pressure is decreased it appears an opposite behavior, and 
if we analyze the diffusion coefficient along isotherms, a minimum of
the diffusion coefficient can be found at a given 
low density~\cite{net01,net02} 
As a consequence, a minimum in the rotational correlation time
is also expected.\\

The behavior of water over a wide range of pressures and temperatures
can be summarized in a coherent picture~\cite{err01}. The region in the
phase diagram in which {\it thermodynamic anomalies} occur is entirely
inside the region of {\it kinetic} or {\it dynamical anomalies} which
in its turn is contained in the region of {\it structural anomalies}.
This picture suggests that all the anomalous behavior of
water can be explained on basis of its structure.
In this case, within
 the region of structural anomalies the orientational and 
translational order should  be inter-dependent~\cite{err01}.
In order to check this hypothesis,  here we study the 
behavior of the orientational relaxation and 
diffusion coefficient  of stretched supercooled water.
By comparing these two quantities, as well as by
studying the local structure of water~\cite{hea93}, such as
revealed by the angular distribution of neighbors 
and hydrogen bonds, we will be able to 
understand how the structure affects the mobility.\\

\section{Methods}

We performed molecular dynamics simulations using 216 water molecules
described by the extended simple
point charge (SPC/E) model~\cite{ber87}, in the canonical
ensemble (NVT), in a cubic simulation box using periodic
boundary conditions. The Berendsen method for rescaling the velocities
was applied~\cite{ber84}, the electrostatic interactions were calculated
using the reaction field method~\cite{ste82} with cutoff of 0.79 nm. The
equations of motion were solved using the SHAKE algorithm~\cite{all87,ryc77}
with time steps of 1.0 fs for T $>$ 210 K   and 2.0 fs for T = 210 K.\\

The diffusion coefficient D was calculated from the asymptotic 
slope of the time dependence of the mean square displacement.
The orientational relaxation was analyzed using the rotational 
autocorrelation functions~\cite{all87}:

\begin{equation}
C^{(i)} ({\bf e}) = \langle P_i \left[ {\bf e}(t) \cdot {\bf e}(0) 
                                \right] \rangle 
\end{equation}

where ${\bf e}$ is a chosen unity vector describing the orientation 
of the molecule and $P_i$ is the $i$-th order Legendre Polynomial. 
We restrict ourselves to the analysis of the first two Legendre
Polynomials, using two choices of reference vectors: 
(1) the dipole vector and (2) a vector 
perpendicular to the plane of the molecule. The correlation
functions were fitted to a biexponential decay function~\cite{yeh99}:

\begin{equation}
C = a_0 exp (- b t^2 /2) + a_I exp (- t/\tau^I) 
+ a_{II} exp (- t/\tau^{II})
\end{equation}

but in the most of the cases taking only one exponential yielded a 
reasonable result, and thus only one relaxation time was determined.\\

In order to understand the effect of the structure on the mobility
we also carried out a detailed analysis of the local structure of
water~\cite{hea93}: the distribution of the hydrogen bond angle
(O-H $\cdots$ O) and the distribution of the angle between
water molecules linked to a central water molecule
(that is, the angle O $\cdots$ O $\cdots$ O).\\

\section{Results}

\begin{table}[!h]
  \begin{center}
\caption{Results of the simulations. $\tau_x^{(y)}$ means the orientational
relaxation time obtained by exponential fit of the $y$-th order Legendre
Polynomial correlation using the $x$-th vector (1 = dipole, 2 = 
perpendicular)}
  \begin{tabular}{lcccccc}
    \hline
 T    & $\rho$              &  Diffusion coefficient
& $\tau_1^{(1)}$  & $\tau_1^{(2)}$  & $\tau_2^{(1)}$  & $\tau_2^{(2)}$    \\
 (K)  & ($g \times cm^{-3}$)&    ($\times 10^{-5} cm^2 s^{-1}$)
& (ps)  & (ps)  & (ps)  &  (ps)   \\
    \hline
 280  & 0.875 & 1.281 & 9.19 & 2.30 & 5.99 & 1.87 \\
 280  & 0.900 & 1.261 & 9.46 & 2.44 & 6.27 & 1.87 \\
 280  & 0.925 & 1.234 & 9.12 & 2.29 & 6.04 & 1.83 \\
  260 & 0.875 & 0.531 & 23.47 & 9.47& 15.65 & 4.91 \\
  260 & 0.900  & 0.527& 22.16 & 8.77& 14.51 & 4.60 \\
  260 & 0.925 & 0.500 & 26.93 & 11.83& 18.09 & 6.52 \\
  250 & 0.875 & 0.298 & 47.69 & 22.55& 31.08 & 19.68 \\
  250 & 0.900 & 0.295 & 43.89 & 21.09& 31.47 & 19.08 \\
  250 & 0.950 & 0.281 & 38.73 & 18.47& 25.55 & 15.96 \\
%
  240 & 0.875 & 0.130 & 95.89 & 44.28&  63.39 & 40.19 \\
  240 & 0.900 & 0.105 & 118.7 & 57.27& 81.28   & 54.92   \\
  240 & 0.925 & 0.122 & 103.82 & 49.41& 68.56 & 29.27 \\
  240 & 0.950 & 0.149  & 75.37 & 36.21&  51.92   & 27.48   \\
  240 & 0.975 & 0.1783 & 66.06 & 31.83& 46.33  & 28.61  \\
  240 & 1.000  & 0.2095 & 56.63 & 25.77&  39.74   & 22.84   \\
  240 & 1.125 &  0.2913 & 28.85& 13.14&  19.32   & 2.06   \\
 230 & 0.875 & 0.0601 & 239.8 & 114.0& 161.8 & 103.1 \\
 230 & 0.900 & 0.0435 & 314.3 & 163.3& 210.8 & 159.1 \\
 230 & 0.925 & 0.048 & 238.8 & 119.6& 161.8 & 112.2 \\
 220 & 0.900 & 0.0114 & 623.9  & 383.7& 401.1 & 379.6  \\
 220 &  0.925 & 0.00612 & 1207.8 & 655.9&  761.6 & 647.0 \\
 210 & 0.875 & 0.00373 & 3582 & 961&  2736 & 2959  \\
 210 &  0.925 & 0.00303 & 3441 & 2393&  2458  & 2412  \\
  \hline
  \end{tabular}
  \end{center}
\end{table}

Our  results are summarized  in Table 1.
The presence of a second relaxation time
(the faster one) is found mainly (but not only) at the higher
densities, in agreement with the interpretation that
these modes are linked to a disturbance
of a tetrahedral local structure~\cite{yeh99}. Since 
the contribution of this faster relaxation time
is almost negligible, we show only the slow 
relaxation times in the Table 1.\\

\clearpage

The diffusion coefficient, as illustrated in Figure $1$,
has a minimum at low densities~\cite{net01}.
If we take into account the results at high pressures~\cite{sta99}, a sigmoidal behavior
is clear. The lines of minima and maxima of the diffusion
coefficient are the borderlines of the region of kinetic or dynamic
anomalies, and it can be shown 
that the thermodynamic anomalies occur 
inside the region of dynamic anomalies~\cite{err01,net01}.\\

The minimum in the diffusion coefficient is located at a density
similar to ice I$_h$. It is interesting also to analyze, whether
the orientational relaxation has a similar dependence 
on density.\\

\vspace*{0.8cm}

\includegraphics[height=7cm]{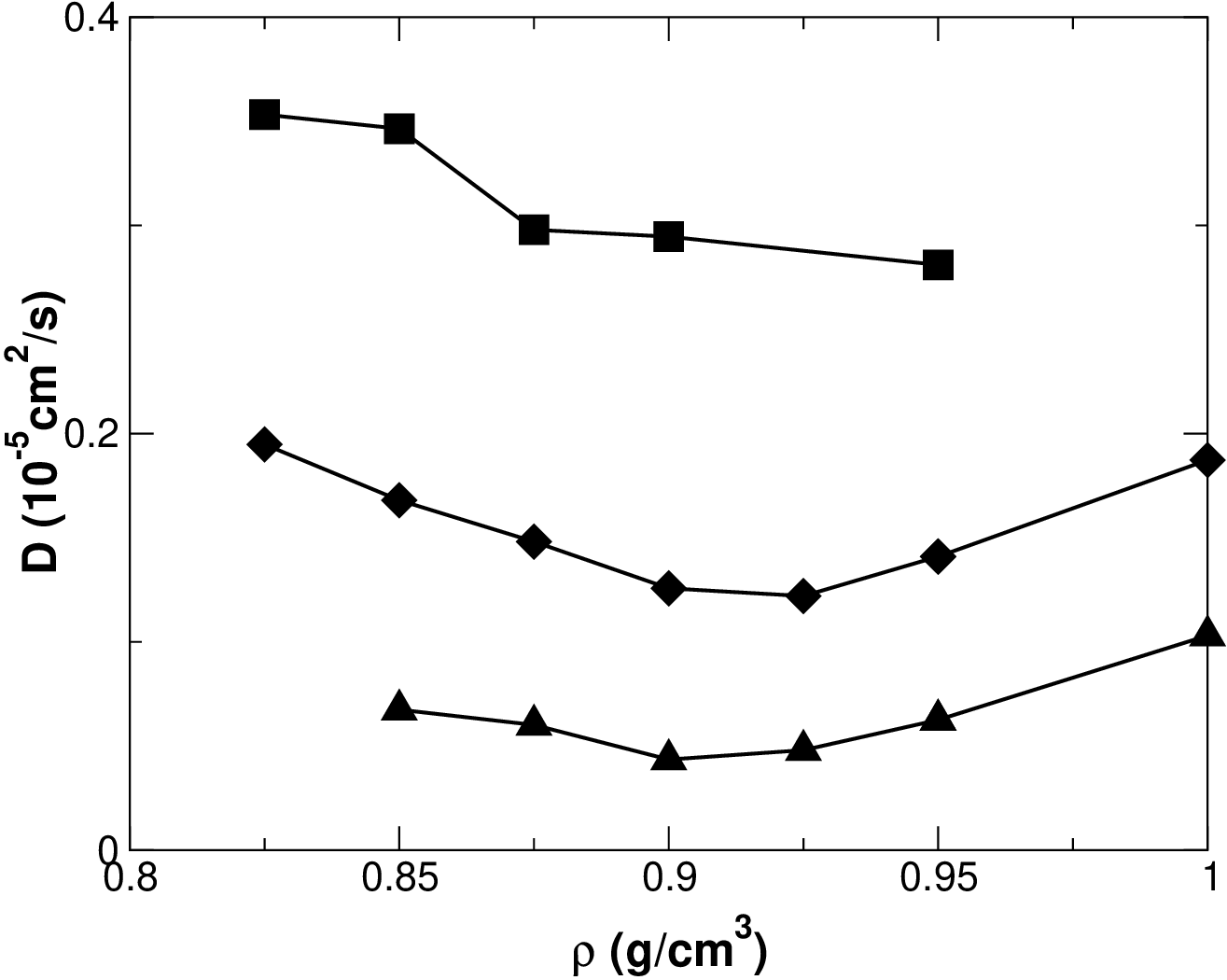}

\vspace*{0.4cm}

\begin{minipage}[c]{.6\textwidth}
{\sf Fig 1. Diffusion Coefficient versus Density along isotherms: 
Detail of the minimum, from top to bottom T = 250 K, 240 K and T = 230 K}

\end{minipage}
\vspace*{0.8cm}

The analysis of the reorientational correlation times show 
peculiar  aspects. First, we can look (see 
Table 1) at the values of the relaxation times for each
chosen vector in order to see some trends. At
 high temperatures, the relaxation times
using the first order Legendre polynomial has nearly three times
the value of the corresponding second order Legendre
polynomial.   This ratio
decays to  about  two or even less at lower temperatures.
This, however, can be attributed to enhancement of the 
 fluctuations at the supercooled region. According to the
Debye rotational diffusion model, this ratio should be
indeed three~\cite{rav97}. The agreement between
our results and the theory is quite striking since
for the mobility of water in its own medium,
the hydrodynamic description should not be valid.\\

The behavior of
the relaxation times with the  density, as can be seen
in Table 1, was also analyzed for  the several vectors
using either first or second order Legendre polynomials. 
At high temperatures, $T=280\;K$ and $T=260\;K$, 
the relaxations times show no specific trend with 
the density.  There is a small 
decrease at low densities, but this decrease
can be attributed to  statistical fluctuations.
At lower temperatures a maximum  is observed  at about the same density as the
translational diffusion coefficients display
their minima~\cite{net01,net02}. This result shows that the
enhancement of the water structure has both influence
in the translational as well as in the rotational
diffusion, as expected.\\

\vspace*{0.8cm}

\includegraphics[height=7cm]{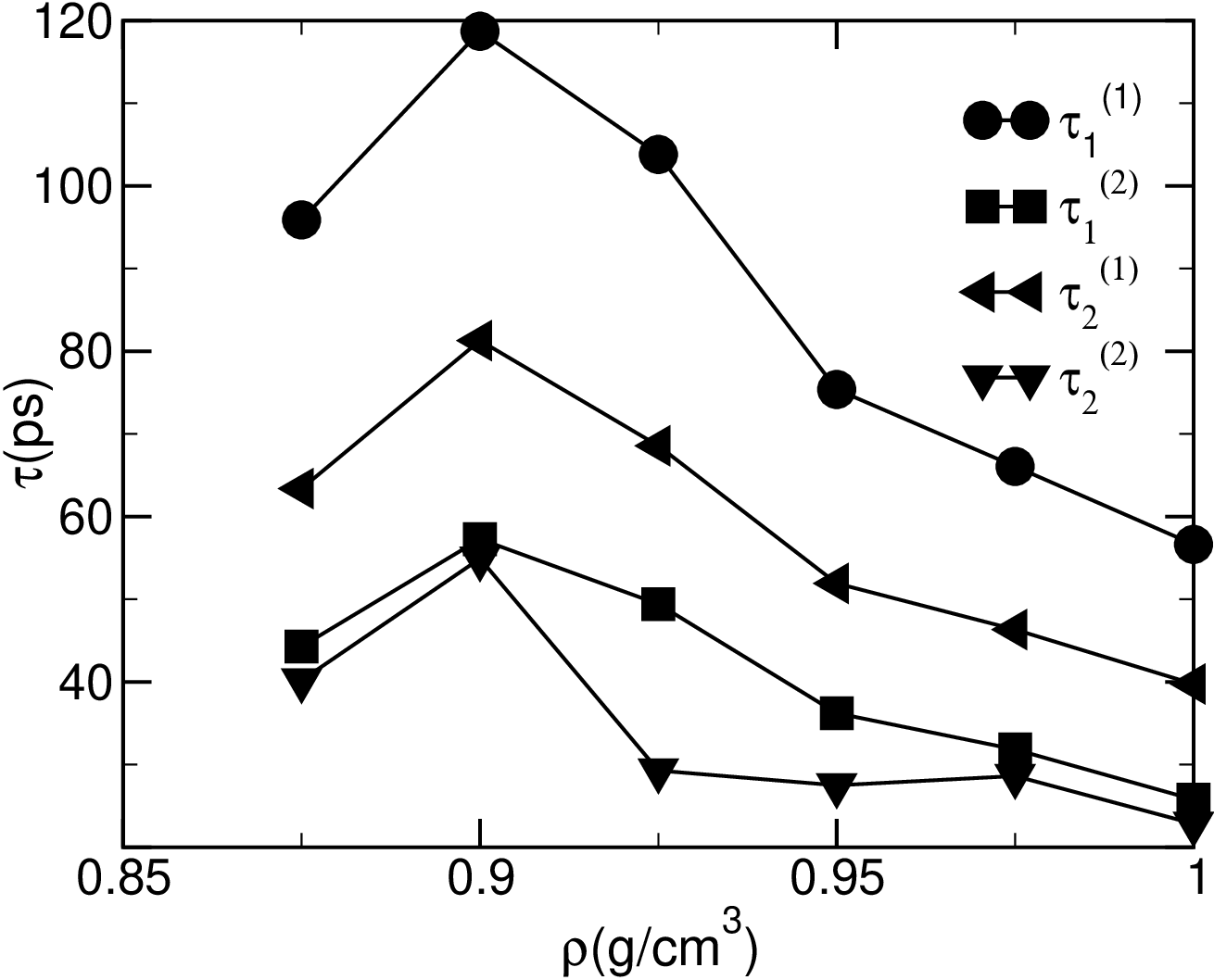}

\vspace*{0.4cm}

\begin{minipage}[c]{.6\textwidth}
{\sf Fig 2. Relaxation times obtained from orientational correlation 
as discussed in the text, for simulations 
at T = 240 K and several densities}
\end{minipage}
\vspace*{0.8cm}

For studying  the mobility of large molecules in
a solvent made of small particles, hydrodynamic arguments
are valid. In that case,  the product
of diffusion coefficient and orientational relaxation time
is a constant, independent of density, viscosity and
temperature~\cite{rav97}.
For studying the mobility of water in its own solution, neither
the continuous description of the solvent nor modeling the
water molecule as compact object are appropriated, so the
hydrodynamic description is expected not to be valid.
Nevertheless, the similarity between  the results for diffusion coefficient 
and relaxation
times  seems to indicate that these two
quantities are related. Therefore we 
computed the product of $D\times\tau$. For all simulations (see Figure 3)
this product remains almost constant what suggests that  the translational
and rotational
mobility are  controlled by a common mechanism.\\

\vspace*{0.8cm}

\includegraphics[height=7cm]{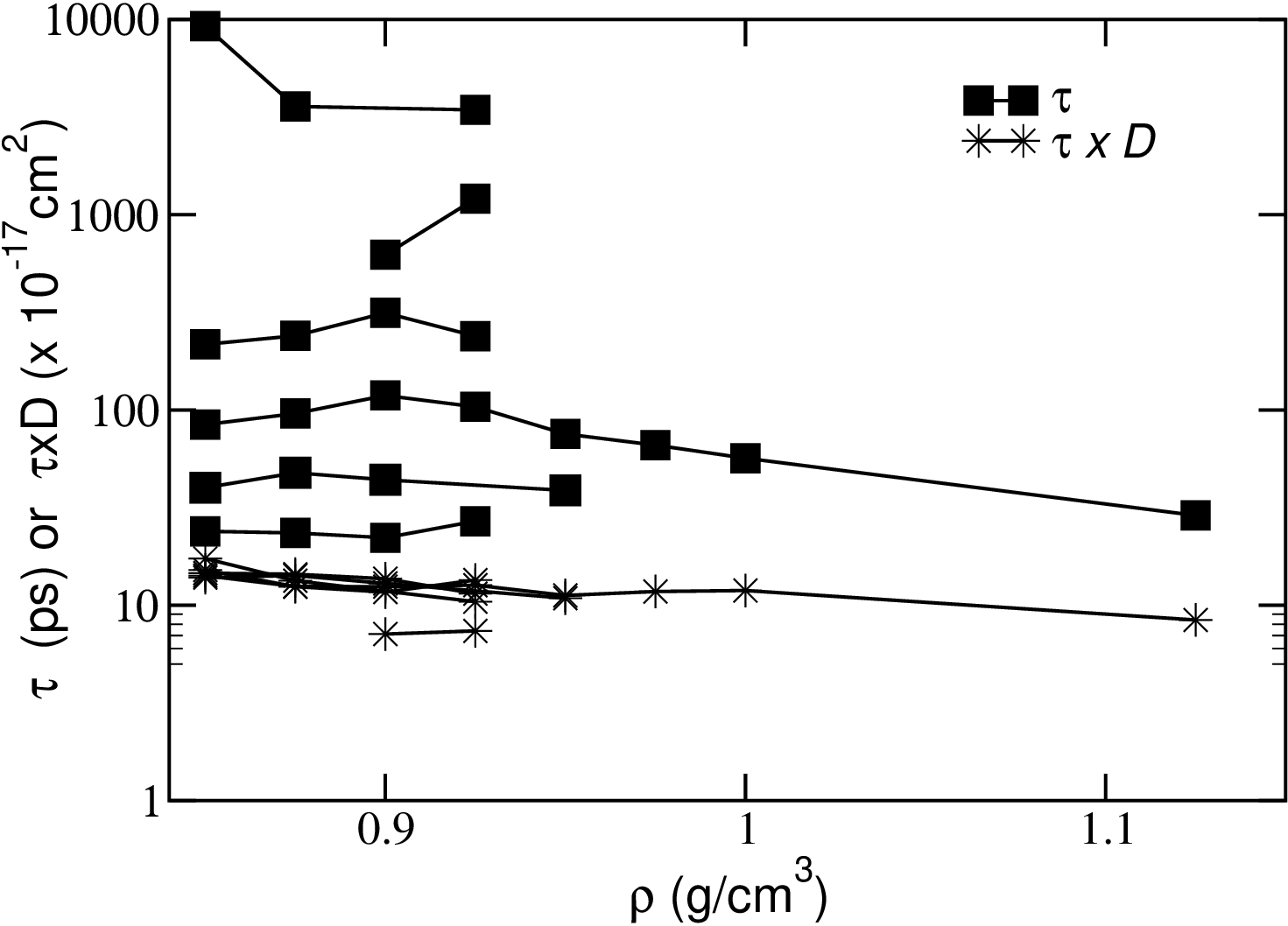}

\vspace*{0.4cm}

\begin{minipage}[c]{.6\textwidth}
{\sf Fig. 3 The product of diffusion coefficient and relaxation time
is nearly constant, independent of the temperature and density}
\end{minipage}
\vspace*{0.8cm}

So if not due to hydrodynamics, how can one explain that
the correlations between $D$ and $\tau$?
Both anomalies in the translational and rotational diffusion are
located inside the region of structural anomalies what
seems to indicate that the structure controls the
mobility.  Even though we have no definitive proof of this 
statement, we have some clues.
A detailed analysis of the local structure of water  reveals 
the enhancement of the tetrahedrality  at low temperatures and ice-like
densities. This can be detected from the  
distribution of the hydrogen bond angle
and the distribution of the angle between neighbors $O \cdots 
O \cdots O$ illustrated  as a function of density for T = 240 K
in Figures  4a,4b  and as a function of temperature for 
density 0.925 g cm$^{-3}$ in  Figures  5a,5b.\\

The first clue comes from the  O $\cdots$ O $\cdots$ O angle
distribution. At low densities, it is centered around the tetrahedral angle,
(109.5$^o$) confirming the enhancement of the ice-like structure.
However, a small peak at angle 60 $^o$ is found. This peak
 increases  slightly with increasing
densities until the density ($\rho_M$ = 1.125 g/cm$^3$)
and  it is  related to the presence of a fifth neighbor molecule
in the first coordination shell. At higher densities, a shoulder
also appears around 90 $^o$, that may be related to the 
appearance of a 6th neighbor.
Therefore, at very high densities, the O $\cdots$ O $\cdots$ O
distribution looks very different with strong distortions in
the structure.\\ 

The second clue, comes from the hydrogen bond angle distribution.
The effect of the density is negligible: at low densities all 
curves almost collapse in a single curve with only one 
peak at low angles, indicating a rigid structure. Only
at high densities the distortions become strong.\\

\vspace*{0.8cm}
\includegraphics[height=6.5cm,width=7.5cm]{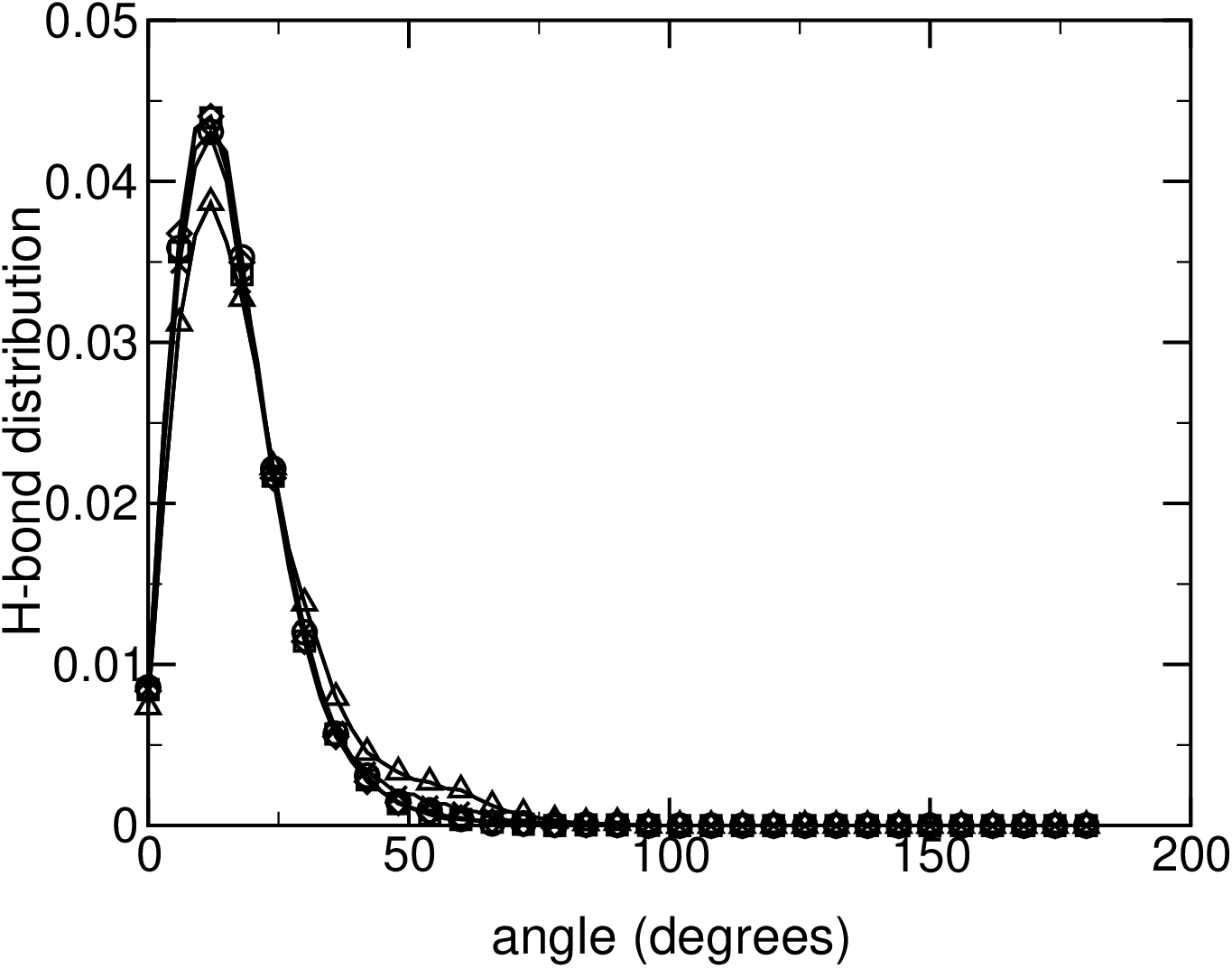}
\hspace*{0.5cm}
\includegraphics[height=6.5cm,width=7.5cm]{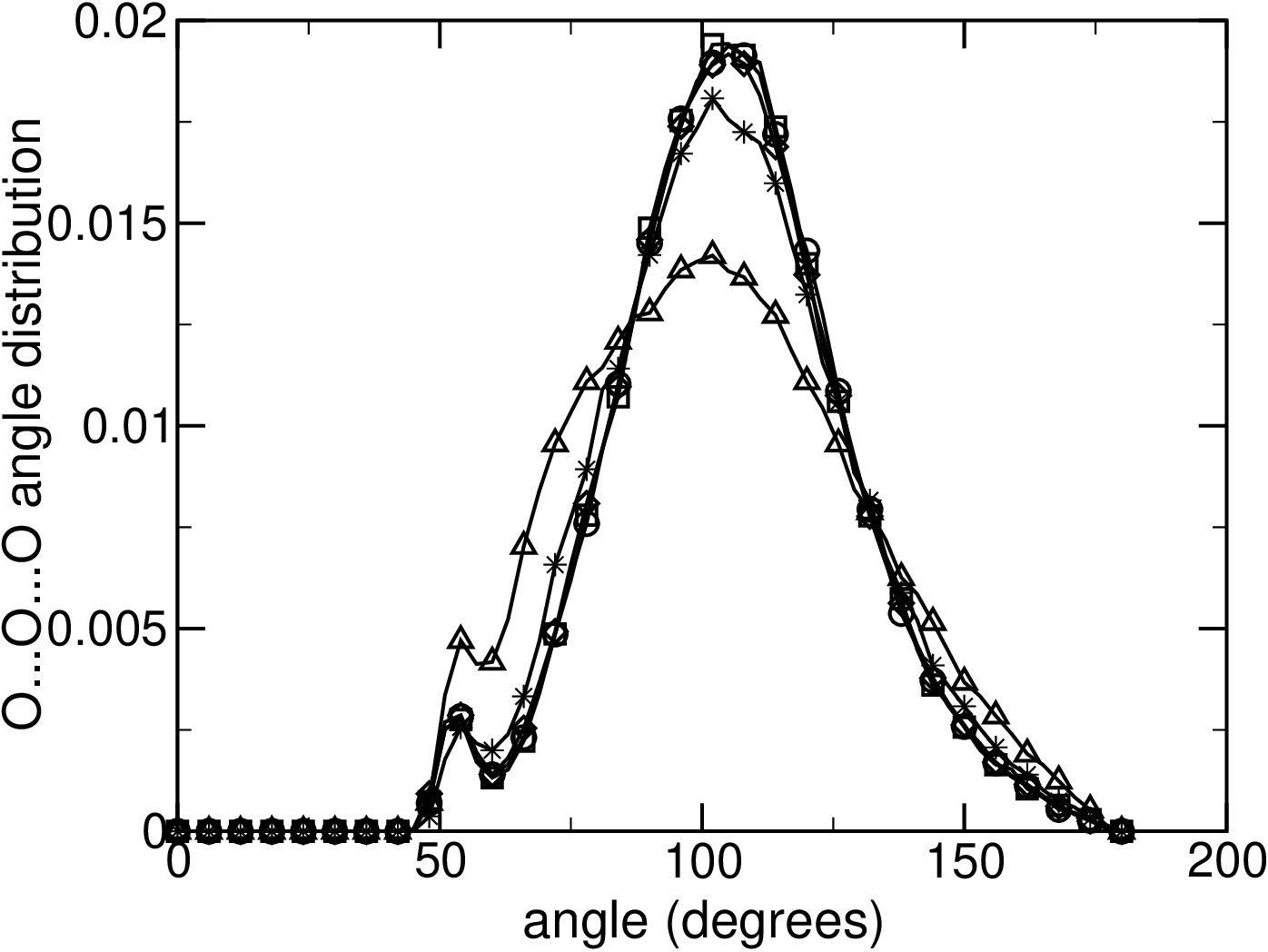}
\vspace*{0.4cm}

\begin{minipage}[c]{.9\textwidth}
{\sf Fig. 4 (a) Hydrogen bond angle distribution  and (b)  neighbors
$O \cdots O \cdots O$ angle distribution  for several densities
at the temperature T = 240 K. Shown are the results for 
$\rho$ = 0.875 g cm$^{-3}$ (O), 
0.900 ($\square$), 0.925 ($\diamond$), 
1.000 ($\star$) and 1.125 g cm$^{-3}$ ($\triangle$)}
\end{minipage}
\vspace*{0.8cm}

\vspace*{0.8cm}
\includegraphics[height=6.5cm,width=7.5cm]{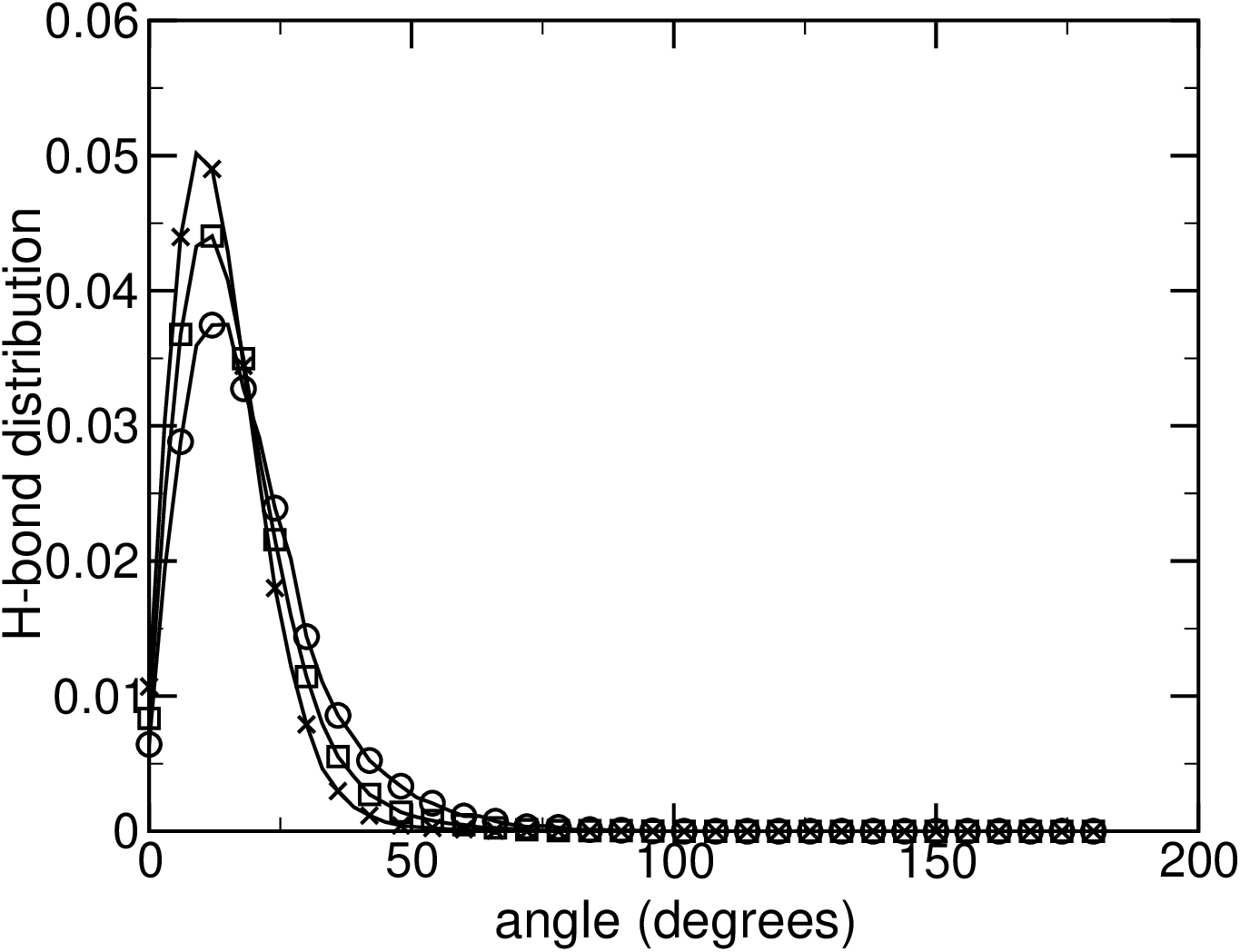}
\hspace*{0.5cm}eurescofig1.eps
\includegraphics[height=6.5cm,width=7.5cm]{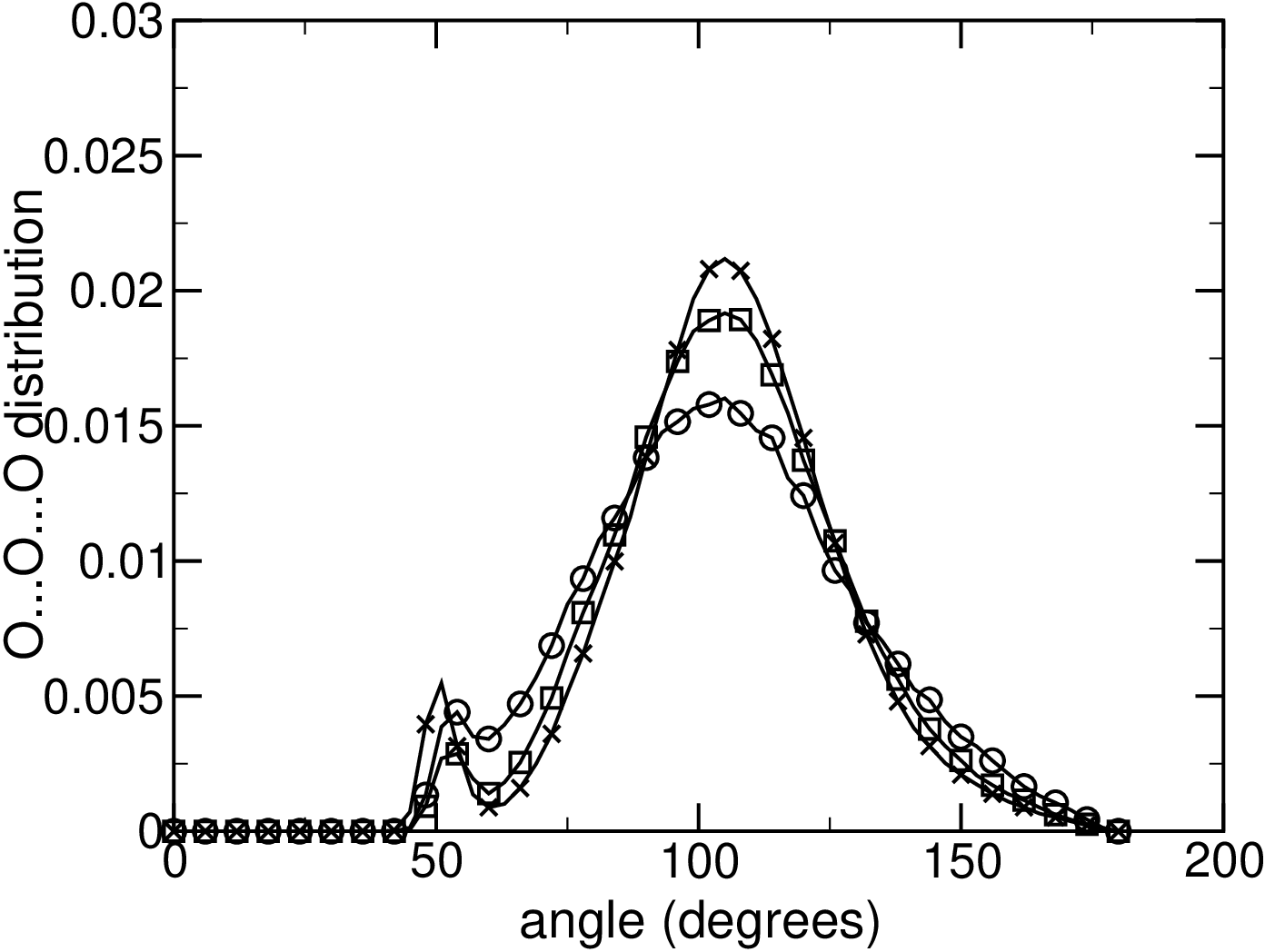}
\vspace*{0.4cm}

\begin{minipage}[c]{.9\textwidth}
{\sf Fig. 5 The same as Fig. 4, but for several temperatures
at fixed $\rho$ = 0.925 g cm$^{-3}$. Shown are the results for 
T = 280 K (O), T = 240 K ($\square$) and T = 210 K ($\star$)} 
\end{minipage}
\vspace*{0.8cm}

As shown in Figure 5a, the  increase of temperature
leads to a broader hydrogen bond distribution, but peaked
in the same value, indicating almost no distortion, apart
from the thermal fluctuation. 
Figure 5b illustrates 
the $O \cdots O \cdots O$ angle distribution. One can observe that the
presence of interstitial water is not strongly affected by
the temperature. The increase of temperature has,
as only effect, a thermal broadening of the distribution.\\

\section{Conclusions}

We have studied the mobility of SPC/E 
water. In contrast with the behavior of a 
normal fluid, the diffusion coefficient of
liquid water increases with the density in a broad range of
thermodynamic conditions, and
exhibits a minimum at the ice-like density
and a  maximum at $\rho_M$. In
compass, the rotational diffusion times increase as
the density is decreased and has a maximum at the
ice-like density. Therefore, in this supercooled stretched region,
the translational 
and rotational mobilities are not independent, but correlated.
The product  of $D\times \tau$ is
a constant in the region of supercooled stretched water.\\

Since the hydrodynamic arguments  are  not valid for describing
the mobility of water in its own medium, we look for
another mechanism for this correlation between $D$ and $\tau$. 
The analysis of the O $\cdots$ O $\cdots$ O  and
hydrogen bond angle
distributions in the stretched region  indicates that the increase in
pressure  
 disturbs the structure by inclusion of interstitial 
molecules but does not cause significative
distortions in the H-bonds. This shows that the increase of
the number of water molecules found in the first coordination shell
and the fraction of water molecules with more than four intact H-bonds
correlates with the increase of the mobility. This suggests that the 
structural defects are directly associated with the anomalous
behavior of translational and rotational diffusion.\\

\end{document}